\documentclass[aps,pre,twocolumn,superscriptaddress,floats,floatfix]{revtex4-2}
\usepackage{amssymb,amsmath,amsthm}
\usepackage{graphicx}
\usepackage{epstopdf}
\usepackage{color}
\usepackage{subfigure}
\usepackage{epsfig}
\usepackage{rotating}
\usepackage{mwe}
\usepackage{float}
\usepackage{dcolumn,bm}
\usepackage{verbatim}
\usepackage{hyperref}
\usepackage{pgf,tikz}
\usetikzlibrary{calc}
\usetikzlibrary{arrows.meta, angles, quotes}
\usepackage{makecell}
\usetikzlibrary{3d}
\usepackage[normalem]{ulem}
\usepackage{xcolor}

\definecolor{Darkred}{RGB}{180, 0, 0}
\definecolor{Darkblue}{RGB}{26,68, 171}
\hypersetup{
  colorlinks = true,     
  urlcolor   = Darkblue, 
  linkcolor  = Darkblue, 
  citecolor  = Darkred   
}

\begin{document}

\begin{abstract}
Magnetars, highly magnetized neutron stars, host superconducting and superfluid phases. We develop a minimal model that captures the interplay between neutron superfluidity, proton superconductivity, and electromagnetic fields using the Gross-Pitaevskii-Poisson, Ginzburg-Landau, and Maxwell equations. Our numerical simulations show that strong rotation enhances the net magnetic field inside the magnetar, suppresses superconductivity there, and amplifies the field near the surface. We explain this by a theory that makes testable predictions, including gravitational-wave signatures.

\end{abstract}
\title{Dynamics of Superfluid-Superconducting Magnetars: Magnetic Field Evolution and Gravitational Waves}

\author{Sanjay Shukla}
\email{shuklasanjay771@gmail.com}
\affiliation{Department of Applied Physics and Science Education, Eindhoven University of Technology, 5600 MB Eindhoven, The Netherlands}
\author{Rahul Pandit}
\email{rahul@iisc.ac.in}
\affiliation{Centre for Condensed Matter Theory, Department of Physics, Indian Institute of Science, Bangalore 560012, India}

\maketitle

\par{\textit{Introduction:}}
The neutrons and protons inside neutron stars have been hypothesized to form superfluids and superconductors, respectively, as was first suggested by Migdal~\cite{MIGDAL_1959} and later developed by Baym, Pethick, and Pines~\cite{Gordon_1969}. Superfluidity and superconductivity in these dense celestial bodies lead to several fascinating phenomena. For instance, they provide an explanation for glitches in the rotation rate of pulsars~\cite{Radhakrishnan_1969,Boynton_1969,Watanabe_2017}, which are magnetized neutron stars, as shown explicitly in recent studies~\cite{verma2022rotating,shukla2024neutron}.

As a neutron star rotates, the neutron superfluid responds by forming quantum vortices, with the quantum of circulation $\mathcal{K} = h/m$~\cite{Onsager_1949}, while the proton superconductor develops flux tubes, with the quantum of magnetic flux $\Phi_0=h/q$~\cite{Loder_2008}. If the rotation frequency of the neutron star is very high, material from the outer layers is ejected, so the diameter of the star is reduced. This shrinkage results in an increase in its rotation speed~\cite{Ryan_2010}, and a consequent generation of closely spaced flux tubes~\cite{babaev2014rotational,shukla2024neutron} that yield, in turn, a strong magnetic field. Finally, the neutron star undergoes a transition to a \textit{magnetar}~\cite{Duncan_1992ApJ,Ferrario_2006,turolla2015magnetars,mereghetti2015magnetars,kaspi2017magnetars,esposito2021magnetars}\footnote{It is important to note that, after this increase in the rotational speed, magnetars quickly lose a large fraction of their rotational energy. We have not included this loss of rotational energy. This requires frictional losses and a crust, as in the pulsar model of Ref.~\cite{verma2022rotating}.
We will study this in future work.}. 
If this rotation-induced enhanced magnetic field increases beyond the upper critical magnetic field $H_{c2}$~\cite{tinkham2004introduction}, the superconducting state becomes normal. In a magnetar, it has been suggested that 
$H_{c2}$ is a function of the density of the neutron superfluid; and $H_{c2}$ assumes its minimal value near the center and its maximal value at the star's surface~\cite{Sinha_2015,Sinha_2015_a}~\footnote{In laboratory settings, the value of $H_{c2}$ is uniform throughout the sample, but it depends on the temperature~\cite{tinkham2004introduction,Dunlap_2019}. Therefore, when the applied magnetic field crosses $H_{c2}$, the whole sample goes into the normal state.}. Therefore, large magnetic fields can cross $H_{c2}$ near the core of a magnetar; if so, a non-superconducting void forms near the center, whereas the region near the surface remains superconducting;
if the field near the center is $\lesssim H_{c2}$, then the core region displays reduced superconductivity. 

We show that the Gross-Pitaevskii-Poisson model, for the neutron superfluid inside a neutron star, coupled with the real-time Ginzburg-Landau equations for the proton superconductor~\cite{shukla2024neutron} provides a natural framework for (a) studying this variation of $H_{c2}$ inside a neutron star and (b) the transition to a magnetar as $\Omega$, the rotation speed, increases. We demonstrate that, if $\Omega$ is low, the star center is superconducting and is threaded by flux tubes. As $\Omega$ is enhanced, the flux tubes move towards the surface and create a void with reduced superconductibity near the center. Furthermore, matter is ejected from the outer layer of the neutron star; this leads to clear gravitational-wave signatures. 

\par{\textit{Model and Methods:}} The neutrons in a neutron star form Cooper pairs that condense to form a superfluid Bose-Einstein condensate. This state can be described by a complex macroscopic wavefunction $\psi_n$. Similarly, the protons Cooper pairs lead to the formation of a Type-II superconductor, with a complex wavefunction $\psi_p$~\cite{Gordon_1969}. These neutron and proton subsystems are coupled with the electromagnetic vector potential ${\bf A}$ and the gravitational potential $\Phi$. The whole system is governed by the Hamiltonian $\mathcal{H} \equiv \mathcal{H}_n+\mathcal{H}_p+\mathcal{H}_{EM}+\mathcal{H}_{G}$, where the subscripts $n,\,p,\,EM$, and $G$ label the Hamiltonians for the neutron, proton, electromagnetic, and gravitational parts [see Appendix~\ref{sec:hamiltonian}]; $\mathcal{H}$ leads~\cite{shukla2024neutron} to the following equations for $\psi_n$, $\psi_p$, and ${\bf A}$ [dimensionless forms in Appendix~\ref{sec:dimensionless_eqs}]:
\begin{eqnarray}
    i\hbar\frac{\partial \psi_n}{\partial t} &=& -\frac{\hbar^2}{2m_n}\nabla^2 \psi_n-\mu_n \psi_n + g|\psi_n|^2 \psi_n+m_n\Phi\psi_n\nonumber\\
    &+&i\hbar ({\bf \Omega} \times {\bf r}) \cdot \nabla\psi_n\,, \nonumber\\
    i\hbar  \frac{\partial \psi_p}{\partial t} &=&q\phi_{\rm eff}\psi_p+\frac{1}{2m_p}\bigg(\frac{\hbar}{i}\nabla- q{\bf A}_{\rm eff}\bigg)^2 \psi_p-\mu_p\psi_p \nonumber\\
    &+&\alpha_s |\psi_p|^2\psi_p+m_p \Phi \psi_p \,,\nonumber\\
    \frac{1}{c^2}\frac{\partial^2 {\bf A}}{\partial t^2} &=& \nabla^2 {\bf A}+\nabla \times {\bf B}_{0}+\mathbb{P}\bigg[\frac{q}{m_pc^2\epsilon_0} {\bf J}_{p} \bigg]\,,\nonumber \\
     \qquad \qquad
    \label{eq:GPE_GLE_ME}
\end{eqnarray}
\begin{figure*}[!hbt]
    \centering
    \includegraphics[scale=0.25]{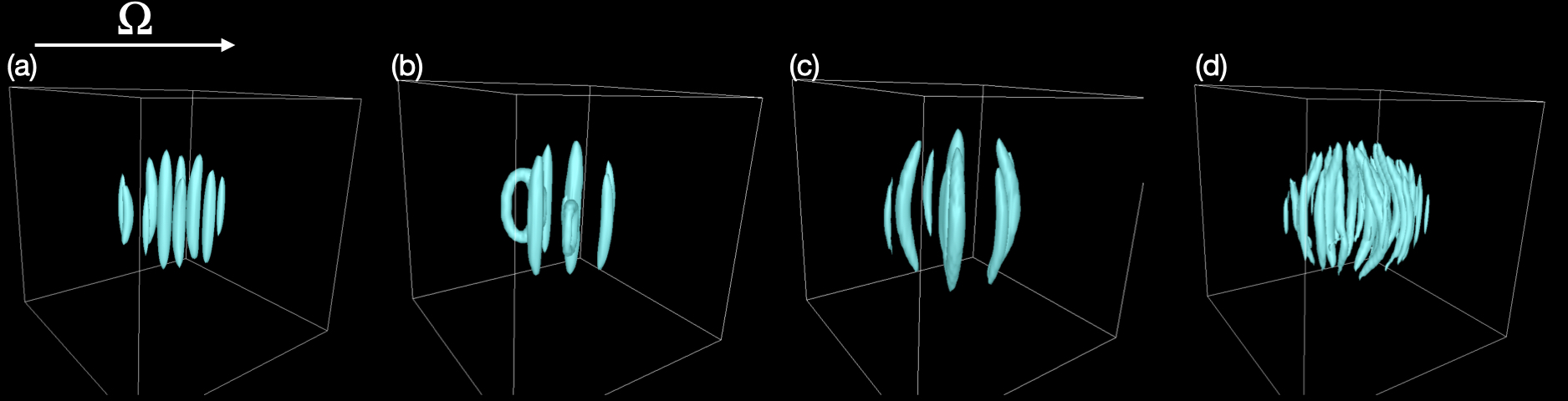}
    \includegraphics[scale=0.25]{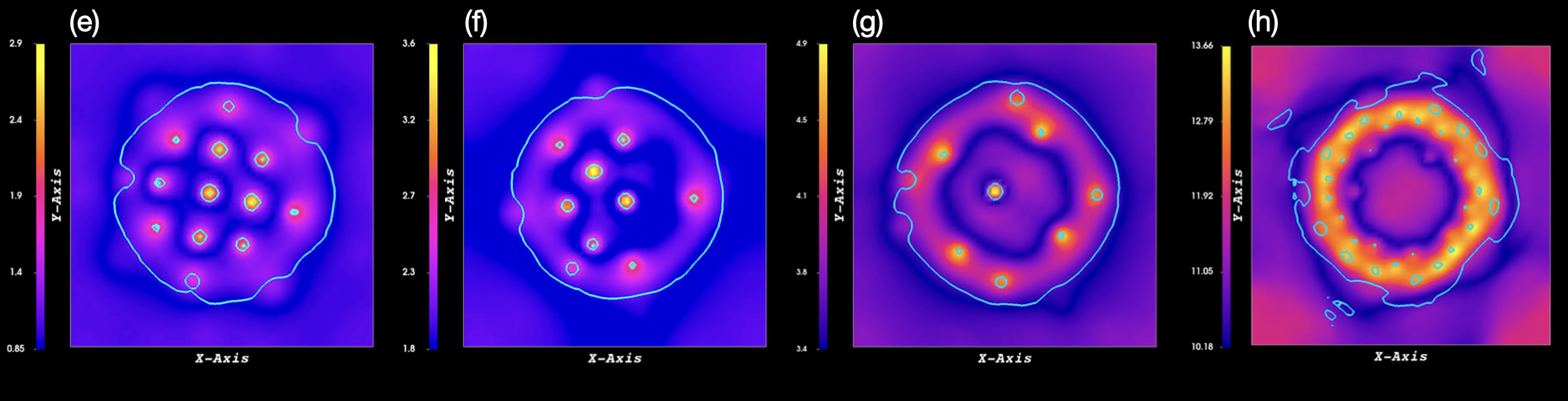}
    \caption{Contour plots of pseudo vorticity ${ \omega}_p = |\nabla \times (\rho {\bf v_n})|$ showing neutron vortices for different rotational speeds (a) $\Omega=0.1\Omega_{\rm ref}$, (b) $\Omega=0.2\Omega_{\rm ref}$, (c) $\Omega=0.4\Omega_{\rm ref}$, and (d) $\Omega=1.2\Omega_{\rm ref}$. Figs. (e) - (h) show the magnetic field distribution $B = |\nabla \times {\bf A}-{\bf B}_{\rm 0}|$ at $z=L/2$ plane with the boundary of the star in cyan color. At these values of $\Omega$, the neutron superfluid does not have vortices.}
    \label{fig:magnetic_field}
\end{figure*}
with the Poisson equations
\begin{eqnarray} 
      \nabla^2 \Phi &=& 4\pi G (m_n|\psi_n|^2 +m_p|\psi_p|^2 )\,, \nonumber\\
       \nabla^2\phi &=& -\frac{1}{\epsilon_0} q|\psi_p|^2\,,
     \label{eq:Poisson}
\end{eqnarray}
where $m_n$ and $m_p$ are the masses of the neutron and proton Cooper pairs, respectively, $g$ and $\alpha_s$ are the self-interaction between neutron Cooper pairs and proton Cooper pairs, respectively, $\mu_n$ and $\mu_p$ are the chemical potentials of neutron and proton Cooper pairs, respectively, $q$ is the charge of the proton Cooper pair, and ${\bf \Omega}$ is the rotational velocity. The effective scalar potential is $\phi_{\rm eff} = \phi-\frac{m_p}{2q}\Omega^2 r^2$, ${\bf A}_{\rm eff} = {\bf A}+\frac{m_p}{q}({\bf \Omega} \times {\bf r})$, ${\bf J}_p = \frac{\hbar}{2i}(\psi_p^*\nabla\psi_p - \psi_p\nabla \psi_p^*) - q {\bf A}_{\rm eff}|\psi_p|^2$, $\bf{B} = \nabla \times {\bf{A}}$, and ${\bf{B}}_0=B_0 \hat{\bf{z}}$ the time-independent mean magnetic field that points in the $\hat{\bf{z}}$ direction. We use the Coulomb gauge $\nabla \cdot {\bf A}=0$, which is maintained by the Helmholtz projector, with the components $\mathbb{P}_{ij}:= \delta_{ij}-\mathcal{F}^{-1} \frac{k_ik_j}{k^2}\mathcal{F}$ and $\mathcal{F}$ the Fourier-transform operator. 

\begin{figure*}[!hbt]
    \centering
    \includegraphics[scale=0.24]{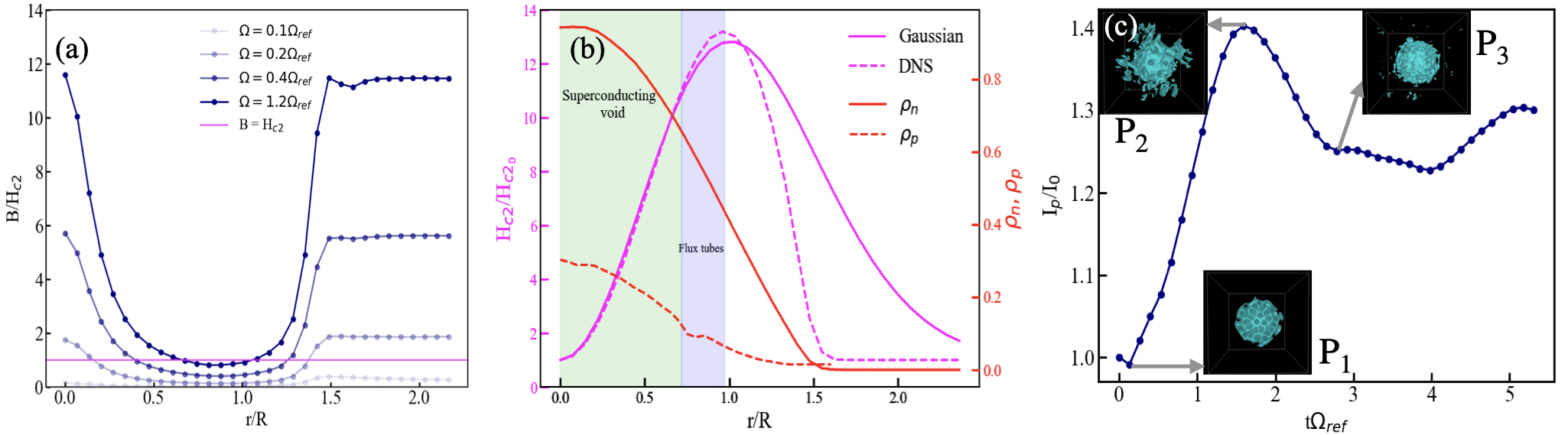}
    \caption{(a) The ratio of the magnetic field and density-dependent upper critical field $B(r)/H_{c2} = |\nabla \times {\bf A}-{\bf B}_{\rm 0}|/H_{c2}$ as a function of the distance $r/R$ for different rotational speeds $\Omega$ in units of the reference frequency $\Omega_{\rm ref}$ [see Appendix~\ref{sec:dimensionless_eqs}]. (b) The plot of the upper critical magnetic field $H_{c2}/H_{c2_0}$ versus the scaled distance $r/R$ at $\Omega = 1.2\Omega_{\rm ref}$ using the Gaussian Ansatz [solid purple curve] and from our DNS [dashed purple curve]; the neutron- and proton-Cooper-pair densities $\rho_n(r)$ and $\rho_p(r)$ are also shown. (c) Plot of the moment of inertia $\mathcal{I}_p/\mathcal{I}_0$ versus the scaled time $t\Omega_{ref}$, using Eq.~\eqref{eq:rgyr} at $\Omega = 1.6\Omega_{\rm ref}$, where $\mathcal{I}_0$ is the moment of inertia at time $t=0$.}
    \label{fig:magnetic_field_highrot}
\end{figure*}

We solve Eqs.~\eqref{eq:GPE_GLE_ME}-\eqref{eq:Poisson} in a cubic domain, with side $L=2\pi$, using pseudospectral direct numerical simulations (DNSs) with $N^3$ collocation points, periodic boundary conditions in all three spatial directions, and the $2/3$-rule for dealiasing, i.e., we set the Fourier modes of all fields to zero, for wave vectors ${\bf{k}}$ with $ |{\bf k}| > k_{max}$ and $k_{max} = [N/3]$~\cite{shukla2024neutron,shukla2024gravity,canuto2006spectral}.  

We use the initial condition $\psi_{n}({\bf{r}},t=0)=\psi_U\exp(\imath \theta({\bf{r}}))$, with $\psi_U$ a positive real constant and $\theta({\bf{r}})$ independent random numbers distributed uniformly on the interval $[0,2\pi]$. For the proton condensate we use the flux-lattice initial condition with $ \psi_{p}({\bf{r}},t=0)= \psi_U\times [\cos(mx)+\imath  \cos(my)]^l$, which is independent of $z$; the positive integer $l$ denotes the multiplicity of a flux tube and $m$ is the number of flux tubes in $-\pi/l \leq x, y \leq\pi/l$. The initial vector potential 
is ${\bf{A}}({\bf{r}},t=0) = {\bf{B}}_{0}\times {\bf{r}}/2$.

\par{\textit{Magnetic field distribution:}} 
Our DNS of Eqs.~\eqref{eq:GPE_GLE_ME}-\eqref{eq:Poisson}, with a small rotational speed $\Omega=0.1\Omega_{ref}$, yields a gravitationally collapsed, approximately spherical star, threaded by flux tubes that we depict by cyan isosurface in Fig.~\ref{fig:magnetic_field}(a). As $\Omega$ increases, these isosurfaces are distorted as shown in Figs.~\ref{fig:magnetic_field} (b)-(d). The pseudocolor plots of Figs.~\ref{fig:magnetic_field} (e)-(h) [corresponding, respectively, to Figs.~\ref{fig:magnetic_field} (a)-(d)] show $|\nabla \times {\bf A}-B_0|$, in the $z=L/2$ plane; the cyan line shows the projection of the boundary of the star, defined by the contour $|\psi_p({\bf{r}})|^2 =  \rho_{th}/m_p$, where $\rho_{th}$ is the threshold density containing $99\%$ of the total mass of the star. Figures~\ref{fig:magnetic_field} (e)-(h) reveal that the magnetic field is concentrated inside the flux tubes, in the interior of the star boundary; such
flux tubes are characteristic of Type II superconductivity within neutron stars and magnetars. As we increase $\Omega$, we observe a reduction in the number of flux tubes in the interior of the star; at large enough $\Omega$, the flux tubes migrate towards the surface [Fig.~\ref{fig:magnetic_field}(h)]. At the same time, the flux tubes near the central region begin to curve [Fig.~\ref{fig:magnetic_field}(d)], creating a roughly spherical region with depleted superconductivity.

Given that the proton Cooper pairs are charged, a part of their response to the rotation of the star is a solid-body rotation $\tfrac{m_p}{2}\Omega^2r^2$, which enters via the square of ${\bf A}_{eff} = {\bf A}+\tfrac{m_p}{q}({\bf \Omega}\times {\bf r})$ in Eq.~\eqref{eq:GPE_GLE_ME}. This rotation leads to  a centrifugal force on the  proton Cooper pairs and an enhanced number of flux tubes, which are pushed away from the star's center and toward its boundary [Fig.~\ref{fig:magnetic_field} (h)]. 
Each flux tube contributes a quantum of magnetic flux to the star, which makes up the total magnetic field. The rotation-induced enhancement of flux tubes causes an increase in the magnitude of the mean magnetic field, as we show via the plots of ratio $B/H_{c2}$ versus $r/R$  in Fig.~\ref{fig:magnetic_field_highrot} (a), where $r$ is the distance from the center of the star and $R$ the radius of gyration:
\begin{eqnarray}
    R = \sqrt{\frac{\mathcal{I}}{\int \rho(r) d{\bf r}}}\,;\;\;  \mathcal{I} \equiv \int \rho(r)r^2 d{\bf r}\,,
    \label{eq:rgyr}
\end{eqnarray}
with $\mathcal{I}$ the moment of inertia. As $B$ increases, superconductivity is depleted. It does not suffice merely to compare $B$ with the upper critical field $H_{c2}$, because the latter also depends on the neutron-Cooper-pair density $\rho_n = |\psi_n({\bf{r}})|^2$. To compute the precise spatial dependence of proton- and neutron-Cooper-pair densities, we carry out a DNS of Eqs.~\eqref{eq:GPE_GLE_ME}-\eqref{eq:Poisson}. This yields the plots, versus $r/R$, of (a) $B/H_{c2}$, for different values of $\Omega$, in Fig.~\ref{fig:magnetic_field_highrot} (a), and (b) $\rho_n$ (full red curve) and $\rho_p$ (dashed red curve) in Fig.~\ref{fig:magnetic_field_highrot} (b). In Fig.~\ref{fig:magnetic_field_highrot} (b) we also show plots of $H_{c2}/H_{c2_{0}}$ (full purple curve, from our DNS, and the dashed purple curve, from a Gaussian approximation that we describe below), where $H_{c2_0} \equiv \tfrac{\Phi_0}{ \xi_p^2}$. We observe that $H_{c2_0}$ is low near the center and assumes its maximal value at the star's boundary. 

As $\Omega$ increases, the flux tubes migrate towards the boundary of the star, which leads to the depletion of superconductivity near the central region. We do not observe a complete vanishing of the density of proton Cooper pairs $\rho_p$, in Fig.~\ref{fig:magnetic_field_highrot} (b), because neutron and proton Cooper pair densities are coupled through the Poisson equation~\eqref{eq:Poisson}. The neutron Cooper-pair density enters into Eq.~\eqref{eq:GPE_GLE_ME} through the gravitational potential $\Phi$; this prevents $\rho_p$ from vanishing completely in Fig.~\ref{fig:magnetic_field_highrot} (b) in the central region $0\leq r/R \lesssim 0.6$. We show, in Fig.~\ref{fig:rho_p} in the Appendix~\ref{sec:sup_density}, that the suppression of $\rho_p$ near this central region increases with increasing $\Omega$. This rotation-induced suppression of $\rho_p$ requires the full set of Eqs.~\eqref{eq:GPE_GLE_ME}-\eqref{eq:Poisson} and has not been obtained by earlier static models~\cite{Sinha_2015,Sinha_2015_a}.

Our model~\eqref{eq:GPE_GLE_ME}-\eqref{eq:Poisson} is also well-suited to uncover the ejection of matter from the star, as we increase $\Omega$; see, e.g., the
ragged edge of the $\rho_n = 99\%$ contour in Fig.~\ref{fig:magnetic_field} (h).
This ejection is observed in magnetars~\cite{rea2010magnetar,turolla2015magnetars,mereghetti2015magnetars,kaspi2017magnetars,esposito2021magnetars}. To illustrate this expulsion of matter, we plot the scaled moment of inertia  $\mathcal{I}(t)/\mathcal{I}(t=0)$ versus the time $t$ in  Fig.~\ref{fig:magnetic_field_highrot}(c).  As a result of the high rotation speed, matter from the surface starts escaping, and the moment of inertia increases from phase ${\rm P}_1$ to ${\rm P}_2$ as we show in Fig.~\ref{fig:magnetic_field_highrot}(c). When the matter leaves the surface completely, the moment of inertia decreases from phase ${\rm P}_2$ to phase ${\rm P}_3$.

\par{\textit{Upper critical magnetic field:}}
The dependence of $B$ and $\rho_p$ on $r/R$ and the depletion of superconductivity near the core suggests that $H_{c2}$ also varies in space, principally because it is a function of the spatially varying neutron density $\rho_n$ [Fig.~\ref{fig:magnetic_field_highrot}]. We can derive an approximate functional form  for $H_{c2}$ as follows: We first seek the stationary-state solutions of the GL part of Eq.~\eqref{eq:GPE_GLE_ME}, with $\Omega=0$, i.e.,
\begin{eqnarray}
    \frac{1}{2m_p}\bigg(\frac{\hbar}{i}\nabla- q{\bf A}\bigg)^2 \psi_p-\mu_p\psi_p
    +\alpha_s |\psi_p|^2\psi_p+m_p \Phi \psi_p\nonumber \\
    +q\phi_{\rm eff}\psi_p=0\,.\nonumber\\
    \label{eq:GLE_equi}
\end{eqnarray}
The time-independent mean magnetic field ${\bf{B}}_0=B_0 \hat{\bf{z}}$, so the initial vector potential is  ${\bf A} = B_0x\hat{{\bf{y}}}$, and the superconducting order parameter depends only on $x$, i.e., $\psi_p = \psi_p(x)$, hence Eq.~\eqref{eq:GLE_equi} becomes
\begin{eqnarray}
    -\frac{\hbar^2}{2m_p}\partial_x^2 \psi_p + \frac{q^2B_{0}^2x^2}{2m_p}\psi_p -\mu_p\psi_p +\alpha_s|\psi_p|^2\psi_p\nonumber \\
    + m_p \Phi \psi_p +q\phi_{\rm eff}\psi_p=0\,;
    \label{eq:non_GLE}
\end{eqnarray}
if we neglect all terms that are explicitly nonlinear in $\psi_p$ [see Appendix~\ref{sec:linear_RTGLE}], we get the Schr\"odinger equation for a harmonic oscillator. From this linearised equation, we obtain the upper-critical field as the solution for the ground-state energy [see Appendix~\ref{sec:linear_RTGLE} for details]:
\begin{eqnarray}
    H^G_{c2}(r) = \frac{\Phi_0}{\xi_p^2} \bigg(1-\frac{1}{2\alpha \beta}\frac{\xi_p^2}{\xi_n^2}\Phi(r) \bigg)\,,
    \label{eq:Hc2_Phi}
\end{eqnarray}
where $\Phi_0=\hbar/q$, and $\Phi$ is the gravitational potential from the Poisson equation~\eqref{eq:Poisson} for the spherical distribution of neutron Cooper pairs [see Appendix~\ref{sec:linear_RTGLE}]. Other parameters like $\xi_n$, $\xi_p$, $\alpha$, and $\beta$ are given in Appendix~\ref{sec:dimensionless_eqs}. The density distribution of neutrons is given by the Gaussian \textit{Ansatz} [Eq.~\eqref{eq:Gaussian_density} in Appendix~\ref{sec:linear_RTGLE}]; the  superscript $G$ refers to the Gaussian \textit{Ansatz}. We show $ H^G_{c2}(r)$ by the solid purple line in Fig.~\ref{fig:magnetic_field_highrot}(b); $H^G_{c2}$ increases with $r/R$, starting from a small value near the core and reaching its maximum near the star's boundary. Thus, the rotation-enhanced magnetic field $B$ crosses $H^G_{c2}$ in the central region of the star, so superconductivity is depleted there. We also calculate $H_{c2}$ from our DNS [dashed purple curve Fig.~\ref{fig:magnetic_field_highrot}(b)]; we see that $H_{c2}$ agrees well
with $H^G_{c2}$ up until the boundary of the star. The Gaussian-\textit{Ansatz} density profile $\rho_n^G$ decreases gradually beyond $r/R=1$, so $H^G_{c2}$ declines slowly beyond the boundary of the star; in contrast, $H_{c2}$ from our DNS exhibits a more rapid decline after this boundary. However, this drop is not exceedingly sharp because of the ejection of matter from the boundary at large $\Omega$.

\par{\textit{Gravitational-wave (GW) signatures:}} To characterize mass ejection from our model magnetar, it is natural to look for gravitational-wave signatures of the type that are used to investigate neutron-star binaries~\cite{tsokaros2024maskingequationstateeffects}. At the level of a first approximation, we can follow the treatment of Ref.~\cite{Lee_1990ApJ} that evaluates the gravitational radiation from Newtonian self-gravitating systems, with mild internal gravity and slow internal motions. In this limit, the gravitational radiation can be measured as the perturbation in the linearized spacetime metric $g_{\mu\nu} = \eta_{\mu\nu}+h_{\mu\nu}$; the components of the perturbations are, in the slow-motion approximation~\cite{Lee_1990ApJ,Vicent_2007_PhysRevD},
\begin{eqnarray}
    h_{ij} = \frac{G}{c^4}\frac{2}{D}\frac{d^2\mathfrak{I}_{ij}}{dt^2}\,,
    \label{eq:GW_strain}
\end{eqnarray}
where $c$ is the speed of light, $D$ is the distance of the source from the observer, and $\mathfrak{I}_{ij}$ are the components of quadrupole moment
\begin{eqnarray}
    \mathfrak{I}_{ij} = \int |\psi_p|^2 (r_i r_j-\frac{1}{3}\delta_{ij}r^2) d^3r\,.
\end{eqnarray}
\begin{figure}[!hbt]
    \centering
    \includegraphics[scale=0.17]{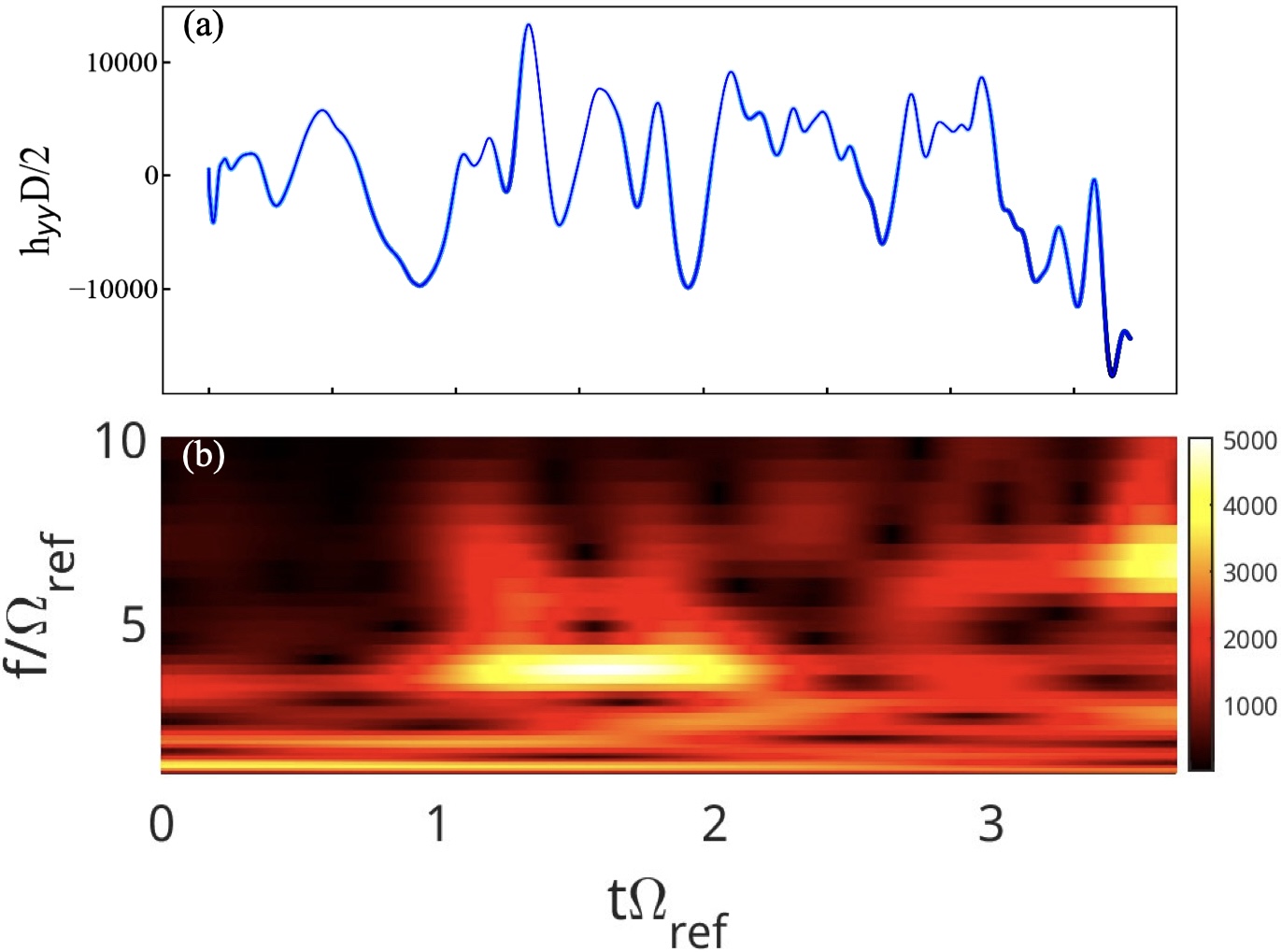}
    \caption{(a) Plot of the gravitational wave strain, from Eq.~\eqref{eq:GW_strain}, versus the scaled time $t\Omega_{ref}$. (b) Pseudocolour plot of
    of the modulus of the continuous wavelet transform (cwt) of $h$ (see text), in the $t\Omega_{ref}$ and scaled frequency $f/\Omega_{ref}$ space,
    showing a shift in frequency with time as the star rotates and flux tubes migrate towards the surface.}
    \label{fig:GW}
\end{figure}
In Fig.~\ref{fig:GW}(a) we plot the gravitational-wave strain $Dh_{xx}/2$ [called $+$ polarisation~\cite{Vicent_2007_PhysRevD}] versus the scaled time $t/\tau$ [the reference time $\tau$ is given in Appendix~\ref{sec:dimensionless_eqs}], which shows distinct speed-ups near $t\Omega_{ref} \simeq 1$ and $3 \lesssim t\Omega_{ref}$. To quantify this,  we perform a continuous wavelet transform (cwt) of the complex array $h=h_{xx}+ih_{yy}$. In Fig.~\ref{fig:GW}(b) we present a pseudocolor plot of the modulus of this cwt of $h$ in the $t\Omega_{ref}$ and scaled frequency $f/\Omega_{ref}$ space. The speed-ups in Fig.~\ref{fig:GW}(a) are associated with the high-intensity yellow regions in Fig.~\ref{fig:GW}(b). Note that the field $B$ also increases with the rotation rate $\Omega$~\footnote{A recent general-relativisitic magnetohydrodynamic (GRMHD) study~\cite{tsokaros2024maskingequationstateeffects} of magnetized neutron stars has also found frequency shifts can be associated with an increase in the magnetic field; however, Ref.~\cite{tsokaros2024maskingequationstateeffects}
does not include superconductivity and superfluidity as we do.}.

\par{\textit{Conclusions:}} Magnetars, the strongest magnets in the known universe, have extreme magnetic fields, densities, and gravity, which pose several challenges for theory, numerical simulations, and experiments~\cite{turolla2015magnetars,mereghetti2015magnetars,Ferrario_2006}. Magnetars, engendered by neutron stars, involve complicated dynamo processes that amplify the magnetic field. High rotation and convection~\cite{Duncan_1992ApJ}, magnetic-flux conservation~\cite{Ferrario_2006}, and superconducting-flux-tube generation away from the magnetar core~\cite{Sinha_2015,Sinha_2015_a} play important roles here. Any self-consistent study of the evolution of the magnetic field in magnetars should consider all the aforementioned process. This is not an easy task. Another major issue is that the observations place very few constraints on the nature and strength of the magnetic field inside magnetars.

The model we have developed addresses a major part of this complicated dynamo process through the rotation-induced enhancement of magnetic flux tubes. In particular, it connects the enhancement of the magnetic field with the rotation, a key observation. Our study goes well beyond earlier theoretical studies by including the spatiotemporal evolution of neutron-superfluid and proton-superconducting states, and the electromagnetic fields using Eqs.~\eqref{eq:GPE_GLE_ME}. Apart from this, our model can be extended to incorporate current-current interactions between neutrons and protons, which lead to the induced magnetic field in the neutron-superfluid, thus enhancing magnetic fields inside the magnetars. Even though it is impossible to realize astrophysical time and length scales [Appendix~\ref{sec:dimensionless_eqs}] in any computational study, our work obtains several desiderata for a model magnetar, such as the frequency- and field-dependence of the depletion of superconductivity in the core region. It also yields, for the first time, GW signatures of the ejection of mass from a rotating magnetar; we hope that these will be tested in experiments.

\begin{acknowledgments}
We thank M.-E. Brachet for discussions, the Anusandhan National Research Foundation (ANRF), the Science and Engineering Research Board (SERB), and the National Supercomputing Mission (NSM), India, for support,  and the Supercomputer Education and Research Centre (IISc), for computational resources. 
\end{acknowledgments}

\section{Appendix}
\subsection{Hamiltonian}
\label{sec:hamiltonian}
The total Hamiltonian describing the dynamics of the neutron-superfluid and proton-superconductor subsystems coupled with the vector potential ${\bf A}$ and gravitational potential $\Phi$ is given by $\mathcal{H} = \mathcal{H}_n+\mathcal{H}_p+\mathcal{H}_{\rm EM}+\mathcal{H}_{\rm G}$, where 
\begin{eqnarray}
\mathcal{H}_{\rm n} &=& \frac{\hbar^2}{2m_n}|\nabla \psi_n|^2 +\frac{g}{2} \bigg(|\psi_n|^2-\frac{\mu_n}{g}\bigg)^2 +m_n\Phi|\psi_n|^2 \nonumber \\
&-&\frac{i\hbar}{2}({\bf\Omega} \times {\bf r}) \cdot (\psi_n\nabla \psi_n^* - \psi_n^*\nabla \psi_n)\,,\nonumber \\
\mathcal{H}_{\rm p} &=&  \frac{1}{2m_p}|D_{\bf A}\psi_p|^2 + \frac{\alpha_s}{2}\bigg(|\psi_p|^2-\frac{\mu_p}{\alpha_s}\bigg)^2 +m_p\Phi|\psi_p|^2\nonumber\\
&-&\frac{1}{2}({\bf \Omega} \times {\bf r}) \cdot (\psi_pD_{\bf A} \psi_p^* + \psi_p^*D_{\bf A} \psi_p)\,,\nonumber\\
 \mathcal{H}_{\rm EM} &=& \frac{\epsilon_0}{2}[{\bf E}^2 - ({\bf B}-{\bf B}_{0})^2]\,;
\label{eq:Hamiltonian_np}
\end{eqnarray}
here, ${\bf E} = -\nabla\phi -\frac{\partial {\bf A}}{\partial t}$, and ${\bf B} = \nabla \times {\bf A}$ are the electric and magnetic fields, respectively. The gravitational Hamiltonian is
\begin{eqnarray}
\mathcal{H}_{\rm G} &=& \frac{1}{8\pi G} (\nabla \Phi)^2\,,
\label{eq:Hamiltonian_G}
\end{eqnarray}
where $G$ is Newton's gravitational constant.

\subsection{Non-dimensionalisation}
\label{sec:dimensionless_eqs}
We obtain the dimensionless forms of Eqs.~\eqref{eq:GPE_GLE_ME}-\eqref{eq:Poisson} by using the reference length $L_{\rm ref}$ and velocity scales $V_{\rm ref}$. The scaled position, time, vector potential, and scalar potential are ${\bf x} = L_{\rm ref}{\bf x}', t = \tfrac{L_{\rm ref}}{V_{\rm ref}} t', {\bf \Omega} = \tfrac{V_{\rm ref}}{L_{\rm ref}} {\bf \Omega}', {\bf A} =  \tfrac{H_{\rm c2}L_{\rm ref}}{\kappa} {\bf A}^{'}, {\rm {and}} \,\, \phi = \tfrac{ L^2_{\rm ref}}{\tau} \tfrac{H_{\rm c2}}{\kappa}  \phi^{'}$. Here $H_{\rm c2}$ is the (zero-temperature) upper critical magnetic field of the superconductor, $\kappa = \frac{\lambda}{\xi_p}$ is the London ratio, $\tau = \frac{L_{\rm ref}}{V_{\rm ref}}$, and $\Omega_{ref} =\frac{V_{\rm ref}}{L_{\rm ref}}$. The scales $L_{\rm ref}$ and $V_{\rm ref}$ can be chosen in many different ways. When we deal with the motion of flux tubes and vortices, we chose $L_{\rm ref}\propto \xi_n$ and $V_{\rm ref}\propto c_s$, where $c_s$ is the speed of sound. The size of the vortex core is chosen based on the size of the grid $\xi_n \propto dx$. The non-dimensionalised equations are
\begin{eqnarray}
i\frac{\partial \psi_n}{\partial t} &=& -\alpha \nabla^2 \psi_n+ \beta (|\psi_n|^2-1) \psi_n  + \mathfrak{G} \Phi\psi_n  \nonumber\\ 
&+&i ({\bf \Omega} \times {\bf r}) \cdot \nabla\psi_n \,,\nonumber\\
i\frac{\partial \psi_p}{\partial t}  &=& \frac{L_{ref}^2\phi_{\rm eff}}{\kappa \xi_p^2} \psi_p+\alpha \bigg(\frac{\nabla}{i}-\frac{L_{\rm ref}^2}{\xi_p^2\kappa} {\bf A}_{\rm eff}\bigg) \psi_p +\mathfrak{G}\Phi \psi_p\nonumber\\
&+&\beta \frac{\xi_n^2}{\xi_p^2}(|\psi_p|^2 -1)\psi_p\,, \nonumber \\
\frac{V_{ref}^2}{c^2}\frac{\partial^2 {\bf A}}{\partial t^2}&=&\nabla^2 {\bf A} +\nabla \times {\bf B}_{\rm ext}+\mathbb{P}\bigg[\frac{1}{\kappa} {\bf J}_{p}\bigg]\,,\nonumber\\
 \nabla^2 \Phi &=&|\psi_n|^2 +\frac{n_p}{n_n}|\psi_p|^2\,, \nonumber\\
 \nabla^2\phi &=& -\frac{\beta}{\kappa} \bigg(\frac{c}{c_s}\bigg)^2 |\psi_p|^2\,.
\label{eq:GPE_neutron_ndim}
\end{eqnarray}
The dimensionless current density and effective vector and scalar potentials are, respectively: 
\begin{eqnarray}
{\bf J}_p &=& \frac{1}{2i} (\psi_p^*\nabla \psi_p-\psi_p\nabla\psi_p^* ) - \frac{L_{ref}^2}{\xi_p^2\kappa} {\bf A}_{\rm eff}|\psi_p|^2\,;\nonumber\\
    {\bf A}_{\rm eff}&=&{\bf A}+\frac{\xi_p^2}{L_{\rm ref}^2} \frac{\kappa}{2\alpha} ({\bf \Omega }\times {\bf r})\,;\nonumber\\
    { \phi}_{\rm eff}&=&{\phi}-\frac{\xi_p^2}{L_{\rm ref}^2} \frac{\kappa}{4\alpha} ({ \Omega^2 }{r^2})\,;
\end{eqnarray}
and
\begin{eqnarray}
\alpha&=&\frac{c_s\xi_n}{\sqrt{2}L_{ref}V_{ref}}\,;\;\; \beta=\frac{c_sL_{ref}}{\sqrt{2}\xi_nV_{ref}}\,;\nonumber \\
\mathfrak{G}&=&\frac{L_{ref}^3 2\sqrt{2}\pi Gm_n n_n}{V_{ref}c_s\xi_n}\,.
\end{eqnarray}
The coherence lengths of the neutron and proton Cooper pairs are $\xi_n=\frac{\hbar}{\sqrt{2m_n gn_n}}$ and $\xi_p=\frac{\hbar}{\sqrt{2m_p \alpha_s n_p}}$, respectively, and $\kappa = \tfrac{\lambda_p}{\xi_p}$ is the London ratio. We can write the following product 
\begin{eqnarray}
    \alpha \beta = \frac{1}{2\mathcal{M}^2}\,,
\end{eqnarray}
where $\mathcal{M}$ is the Mach number. By choosing a suitable Mach number and $\alpha$, the value of $\beta$ can be fixed.

The number of flux tubes inside neutron stars is $\simeq 10^{15}$; in our DNSs they are restricted to $\simeq 30$; the ratio of the speed of light to that of sound $\frac{c}{c_s}\simeq 10^6$, which is a challenge for any DNS. The upper critical magnetic field $H_{c2_0}/\Phi_0 \simeq 10^{30}$ inside neutron stars, which results from the tiny radius of the flux tubes. In our DNS $H_{c2_0}/\Phi_0 \simeq 10^{3}$. However, we get the ratio $\tfrac{\xi_n}{\xi_p}=2$, which agrees with the value found in neutron stars~\cite{shukla2024neutron}. Moreover, the London parameter $\kappa$ is greater than $1/\sqrt{2}$, so that we have type-II superconductivity.

\subsection{Superconducting density with rotation}
\label{sec:sup_density}
\begin{figure}[!hbt]
    \centering
    \includegraphics[scale=0.4]{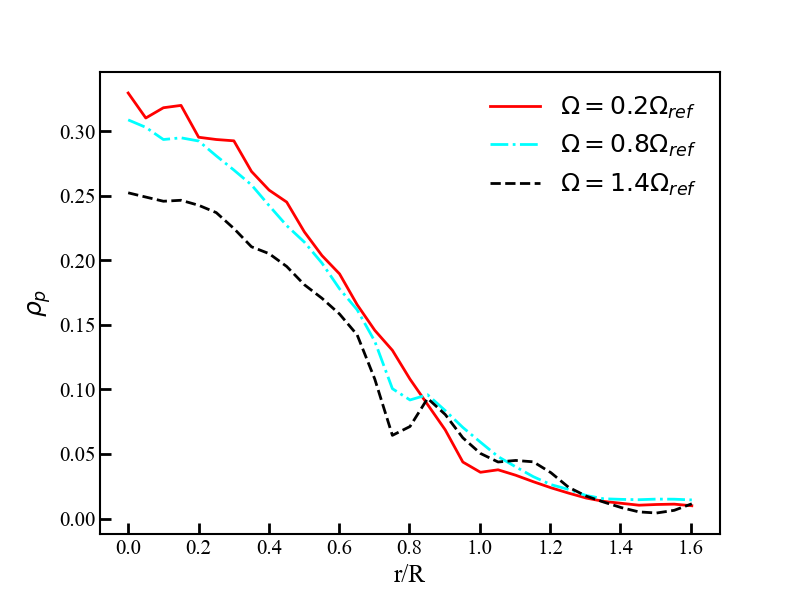}
    \caption{Plots of the proton-Cooper-pair density $\rho_p$ versus the scaled radial distance $r/R$ for different values of the rotation frequency $\Omega$; in the region of the core of the magnetar, $\rho_p$ 
    decreases with increasing $\Omega$.}
    \label{fig:rho_p}
\end{figure}

In Fig.~\ref{fig:rho_p} we present plots of the proton-Cooper-pair density $\rho_p$ versus the scaled radial distance $r/R$ for different values of the rotation frequency $\Omega$; in the region of the core of the magnetar, $\rho_p$  decreases with increasing $\Omega$.

\subsection{Linear Ginzburg-Landau equation}
\label{sec:linear_RTGLE}
For the initial vector potential ${\bf A} = B_0x\hat{{\bf{y}}}$, the superconducting order parameter depends only on $x$, i.e., $\psi_p = \psi_p(x)$, and the Real-time-Ginzburg-Landau Eq.~\eqref{eq:GLE_equi} becomes
\begin{eqnarray}
    -\frac{\hbar^2}{2m_p}\partial_x^2 \psi_p + \frac{q^2B_{0}^2x^2}{2m_p}\psi_p -\mu_p\psi_p +\alpha_s|\psi_p|^2\psi_p\nonumber \\
    + m_p \Phi \psi_p +q\phi_{\rm eff}\psi_p=0\,;
\end{eqnarray}
if we neglect all terms that are explicitly nonlinear in $\psi_p$, we get
\begin{eqnarray}
    -\partial_x^2 \psi_p + \frac{B_{0}^2x^2}{\Phi_0^2}\psi_p =\frac{1}{\xi_p^2} \bigg(1-\frac{1}{2\alpha \beta}\frac{\xi_p^2}{\xi_n^2}\Phi \bigg)\psi_p\,,
    \label{eq:type2_sup_HO}
\end{eqnarray}
where $\Phi_0=\hbar/q$, $\hbar/m=\sqrt{2}c_s\xi_n$, and $\Phi$, the gravitational potential given by Eq.~\eqref{eq:Poisson}, which can be written as 
\begin{eqnarray}
    \nabla^2\Phi \simeq 4\pi Gm_n|\psi_n|^2\,,
    \label{eq:lin_Poisson}
\end{eqnarray}
where we have again neglected terms that are nonlinear in $\psi_p$. For simplicity, we consider the rotation speed to be such that the neutron condensate is devoid of quantum vortices and is given by a spherical density distribution. For such a spherical density distribution, the solution of the gravitational potential, from Eq.~\eqref{eq:lin_Poisson}, can be written as
\begin{eqnarray}
    \Phi(r) &=& -\frac{GM_n(r)}{r}\,,\nonumber \\
    M_n(r) &=& \int_0^r \rho_n(r')4\pi(r')^2 dr'\,, 
    \label{eq:gravt_pot}
\end{eqnarray}
where $M_n(r)$ is the neutron Cooper pair mass contained in a sphere of radius $r$. 

Equation~\eqref{eq:type2_sup_HO} looks like the Schr\"odinger equation for a  harmonic oscillator, but with an additional $r$ dependence because of the gravitational potential $\Phi$. When $\Phi=0$, the ground-state solution of Eq.~\eqref{eq:type2_sup_HO} yields the usual expression for the upper-critical field $H_{c2_0} = \tfrac{\Phi_0}{ \xi_p^2}$. If $\Phi$ depends on $r/R$, the right-hand side (RHS) of Eq.~\eqref{eq:type2_sup_HO} becomes a function of $\rho_n=|\psi_n|^2$, so an analytical solutions cannot be obtained. Therefore, we solve Eq.~\eqref{eq:type2_sup_HO} numerically, for given values of $r$, the approximation~\eqref{eq:gravt_pot}, and the Gaussian-profile \textit{Ansatz}
\begin{eqnarray}
    \rho_n = \rho_0e^{-r^2/R_{ G}}\,,
    \label{eq:Gaussian_density}
\end{eqnarray}
where $R_{G} = \sqrt{2/3} R$ with $R$ radius of gyration~\eqref{eq:rgyr}.

\end{document}